\documentclass[aps,reprint,superscriptaddress,amsmath,amssymb]{revtex4-2}

\usepackage{graphicx}% Include figure files
\usepackage{dcolumn}% Align table columns on decimal point
\usepackage{bm}% bold math
\usepackage{hyperref}% add hypertext capabilities
\usepackage[normalem]{ulem}% strikeout text
\usepackage{xcolor}
\usepackage[utf8]{inputenc}

\begin{document}

\title{Experimental Search for Neutron to Mirror Neutron Oscillations as an Explanation of the Neutron Lifetime Anomaly}%

\author{L.~J.~Broussard}
\altaffiliation[]{broussardlj@ornl.gov}
\affiliation{Oak Ridge National Laboratory, Oak Ridge, TN 37831, USA}

\author{J.~L.~Barrow}
\altaffiliation[]{Now at the Massachusetts Institute of Technology and Tel Aviv University.}
\affiliation{Department of Physics, University of Tennessee, Knoxville, TN 37996, USA}
\author{L.~DeBeer-Schmitt}
\affiliation{Oak Ridge National Laboratory, Oak Ridge, TN 37831, USA}
\author{T.~Dennis}
\affiliation{Department of Physics and Astronomy, East Tennessee State University, Johnson City, TN 37614}
\author{M.~R.~Fitzsimmons}
\affiliation{Oak Ridge National Laboratory, Oak Ridge, TN 37831, USA}
\affiliation{Department of Physics, University of Tennessee, Knoxville, TN 37996, USA}
\author{M.~J.~Frost}
\affiliation{Oak Ridge National Laboratory, Oak Ridge, TN 37831, USA}
\author{C.~E.~Gilbert}
\affiliation{Oak Ridge National Laboratory, Oak Ridge, TN 37831, USA}
\author{F.~M.~Gonzalez}
\affiliation{Oak Ridge National Laboratory, Oak Ridge, TN 37831, USA}
\author{L.~Heilbronn}
\affiliation{Department of Nuclear Engineering, University of Tennessee, Knoxville, TN 37996, USA}
\author{E.~B.~Iverson}
\affiliation{Oak Ridge National Laboratory, Oak Ridge, TN 37831, USA}
\author{A.~Johnston}
\affiliation{Department of Physics, University of Tennessee, Knoxville, TN 37996, USA}
\author{Y.~Kamyshkov}
\affiliation{Department of Physics, University of Tennessee, Knoxville, TN 37996, USA}
\author{M.~Kline}
\affiliation{Department of Physics, The Ohio State University, Columbus, OH 43210, USA}
\author{P.~Lewiz}
\affiliation{Department of Physics, University of Tennessee, Knoxville, TN 37996, USA}
\author{C.~Matteson}
\affiliation{Department of Physics, University of Tennessee, Knoxville, TN 37996, USA}
\author{J.~Ternullo}
\affiliation{Department of Physics, University of Tennessee, Knoxville, TN 37996, USA}
\author{L.~Varriano}
\affiliation{Department of Physics, University of Chicago, Chicago, IL 60637, USA}
\author{S.~Vavra}
\affiliation{Department of Physics, University of Tennessee, Knoxville, TN 37996, USA}

\date{\today}

\begin{abstract}
An unexplained $>4\,\sigma$ discrepancy persists between ``beam'' and ``bottle'' measurements of the neutron lifetime. A new model proposed that conversions of neutrons $n$ into mirror neutrons $n'$, part of a dark mirror sector, can increase the apparent neutron lifetime by $1\%$ via a small mass splitting $\Delta{m}$ between $n$ and $n'$ inside the 4.6\,T magnetic field of the National Institute of Standards and Technology Beam Lifetime experiment. A search for neutron conversions in a 6.6\,T magnetic field was performed at the Spallation Neutron Source which excludes this explanation for the neutron lifetime discrepancy.
\end{abstract}

\maketitle

While observational evidence for dark matter is strong, there is not yet any confirmed direct detection, prompting a shift from the Weakly Interacting Massive Particle (WIMP) paradigm and motivating searches for increasingly diverse candidates~\cite{Battaglieri:2017aum}. The theory of a mirror sector, containing a replica of Standard Model (SM) particles and interactions with only common gravitational interactions with our sector~\cite{Kobzarev:1966qya, Blinnikov:1982eh,Foot:1991bp,Hodges:1993yb,Okun:2006eb}, has garnered renewed attention as a viable hidden sector dark matter candidate~\citep{Berezhiani:2000gw,Ignatiev:2003js,Berezhiani:2003xm,Berezhiani:2003wj,Berezhiani:2005ek,Foot:2014mia}. 
The model is consistent with astrophysical observations for the abundance of dark matter if the exact symmetry between the mirror and ordinary sectors is broken with different vacuum expectation values and with a lower temperature of the mirror sector~\citep{Berezhiani:1995yi, Berezhiani:1995am, Babu:2021mjg}.

The theory of mirror matter has testable consequences if there are new interactions mixing the neutral singlets of SM and SM$'$ such as the neutrino $\nu_R$ with the sterile neutrino $\nu'_R$~\cite{Berezhiani:1995yi, Berezhiani:1995am} or the neutron $n_R$ with the sterile neutron $n'_R$~\cite{Berezhiani:2005hv,Berezhiani:2006je}, leading to oscillation effects between $\nu \leftrightarrow \nu'$ and $n \leftrightarrow n'$. Models with broken symmetry naturally include small mass non-degeneracy between the ordinary and mirror sectors. A small mass splitting between neutrons and mirror neutrons was considered in~\citep{Berezhiani:2009ldq, Berezhiani:2018eds}. Neutron conversions into hidden sectors were also considered in~\cite{Sarrazin:2012sc, Sarrazin:2015sua}. Observation of $n\rightarrow n'$ would have interesting astrophysical implications for neutron stars~\cite{Goldman:2019dbq, Berezhiani:2021src, sym14030518}, in particular if there is a small mass splitting~\cite{Berezhiani:2020zck}, and for extreme energy cosmic rays~\cite{Berezhiani:2006je,Berezhiani:2011da}. Some constraints for neutron disappearance have been obtained with ultracold neutrons ~\cite{Ban:2007tp,Serebrov:2007gw,Serebrov:2008her, Bodek:2009zz, Altarev:2009tg,Berezhiani:2012rq, Berezhiani:2017jkn, nEDM:2020ekj}, with some controversial anomalous signals reported~\citep{Berezhiani:2012rq,Berezhiani:2017jkn}. Future UCN searches are discussed in~\citep{sym14030503, sym14030487}.  An alternative approach has been proposed using the technique of cold neutron regeneration \cite{Berezhiani:2009ldq,Berezhiani:2017azg,Schmidt:2007}. Limits have also been obtained with passing-through-walls experiments~\cite{Sarrazin:2016bsw, Stasser:2020jct, Almazan:2021fvo}.

We report on a first experimental demonstration of the cold neutron regeneration approach at the Spallation Neutron Source (SNS), summarized previously \cite{Broussard:2017yev,Broussard:2019tgw}. This experiment addresses a new theoretical model of non-degenerate mirror dark matter~\cite{Berezhiani:2018eds}, proposed as a potential explanation of the neutron lifetime anomaly--the disagreement between two complementary methods of neutron lifetime measurements. Ultracold neutron (UCN) bottle experiments use magnetic~\citep{Ezhov:2014tna, Pattie:2017vsj} or gravitational~\citep{Serebrov:2004zf, Pichlmaier:2010zz, Steyerl:2012zz, ARZUMANOV201579, Serebrov:2017bzo, Serebrov:2017jvb} traps to measure the rate of neutron disappearance, giving a mean lifetime of $879.4\pm 0.6\,$s~\citep{ParticleDataGroup:2020ssz} ($878.4\pm0.5\,$s including \citep{UCNt:2021pcg}). Cold neutron beam experiments detect the appearance of either protons~\citep{Byrne:1996zz,Nico:2004ie,Yue:2013qrc} or electrons~\citep{Sumi:2021svn} from neutron $\beta$-decay. These beam experiments find a neutron lifetime of $888.1 \pm 2.0\,$s, due primarily to the Beam Lifetime result~\cite{Yue:2013qrc}, which is $>4\, \sigma$ higher than the bottle experiments. 

Although this difference could be attributed to some not yet understood experimental systematic effects in ~\cite{Yue:2013qrc}, it could be the manifestation of new physics. The neutron could decay into dark matter and lower the apparent lifetime determined in ultracold neutron bottles by $\sim$1\%~\citep{Fornal:2018eol,Berezhiani:2018udo}. These proposed decay modes have been constrained by direct searches~\citep{Tang:2018eln,UCNA:2018hup,Klopf:2019afh}. Also, it has been emphasized that consistency of the current neutron $\beta$-decay dataset with Cabibbo-Kobayashi-Maskawa matrix unitarity and superallowed nuclear decays disfavors neutron dark decay models as an explanation for the lifetime discrepancy~\citep{Czarnecki:2018okw,Dubbers:2018kgh,Belfatto:2019swo}.

Alternatively, the non-degenerate mirror dark matter model suggested instead that the Beam Lifetime result~\citep{Nico:2004ie,Yue:2013qrc} is overestimated by $\sim$1\% due to the missed decays of neutrons temporarily being in the $n'$ state~\citep{Berezhiani:2018eds}. In this model, the neutron $n$ mixes with a sterile ``mirror" neutron $n'$ with a non-degenerate  mass $m_{n'}=m_n \pm \Delta{m}$ via an oscillation described by a mixing angle $\theta_0$.
%$\sim200$ neV. 

In Beam Lifetime~\citep{Nico:2004ie, Yue:2013qrc}, a cold neutron beam passes through a $4.6\,$T solenoidal magnet which traps the $\beta$-decay protons, and the rate of the appearance of protons is compared to the neutron flux to determine the lifetime. The model predicts that while neutrons enter and exit the magnet they pass through regions of magnetic potential which compensate the unknown mass splitting $\Delta{m}$, thus enhancing the $n \leftrightarrow n'$ oscillation probability. As neutrons have an enhanced probability to be found in the mirror neutron state inside the magnet, the number of neutrons $\beta$-decaying within the decay volume is reduced. Any decay products of the mirror neutrons are not detectable. The probability to be found as a mirror neutron then reduces after leaving the magnet, where neutrons are detected with a lesser impact on the flux normalization (e.g. see Fig.~3 of ~\citep{Berezhiani:2018eds}). This results in a longer measured lifetime. The probability depends on two parameters: the mass splitting $\Delta{m}$ and the mixing angle $\theta_0$ between the $n$ and $n'$ states in vacuum. Figure 4 of reference~\citep{Berezhiani:2018eds} predicts the range of possible values of $\theta_{0}$ as a function of $\Delta{m}$ which would explain the apparently $1\%$ higher neutron lifetime in the beam experiment than in the bottle experiments.

In \citep{Berezhiani:2018eds}, the probability of the $n\rightarrow{n'}$ transformation was only calculated in the region $\Delta{m} >278\,$neV, above the maximum value of magnetic field used in the Beam Lifetime experiment. This is an adiabatic situation as $\Delta{m}$ is not fully offset by the magnetic field. Negative values of $\Delta{m}$ are not consistent with the neutron lifetime anomaly. The region $\Delta{m} <278\,$neV was not studied due to the intense computational requirements of the calculation in the non-adiabatic regime. To correctly treat smaller positive values of $\Delta{m}$, we must account for the non-adiabatic Landau-Zener (LZ) transitions which occur for one neutron polarization where the magnetic potential $\mu_{n}B(z)$ compensates $\Delta{m}$. The result of the LZ transitions depends on the initial phase at $t=0$ of the oscillating $(n,n')$ system, the shape of the magnetic field profile, and on the neutron velocity $v$ distribution.

The time evolution of the density matrix of the $(n,n')$ system is described by the Hamiltonian 
\begin{equation}
\label{eqn:Hregen}
H=\left(\begin{array}{cc}
V -iW -\Delta{m} \pm \mu_n B(z) & \epsilon{} \\
\epsilon{} & 0
\end{array}\right),\
\end{equation}
where $V>0$ and $W$ are the real and imaginary parts, respectively, of the optical potential of the propagation media (air or the neutron absorber) and $\Delta{m}\equiv m_{n'}-m_{n}>0$. As in \citep{Berezhiani:2018eds} we define tan$2\theta_0 = 2\epsilon{}/\Delta{m}$ which modifies to tan$2\theta_{m} =$ tan$2\theta_0 /(1 \pm \mu B / \Delta{m}$) in a magnetic field. The sign $\pm$ of the $\mu_{n}B$ term corresponds to the state of neutron polarization. The initial state of the density matrix at $t=0$ was assumed to be pure neutron. The evolution of the density matrix for the oscillating $(n,n')$ system in the absorber material used in this experiment are described in~\citep{Kamyshkov:2021kzi}.

We calculated the predicted parameter space consistent with the $(1.0\pm0.2)\%$ effect in the neutron lifetime for positive $\Delta{m}$. This calculation implements the measured magnetic field profile of the Beam Lifetime experiment reported in~\citep{Nico:2004ie}, assuming the magnetic field falls off as a solenoidal field. This extends the work of reference~\citep{Berezhiani:2018eds}, where an idealized model of the Beam Lifetime magnetic field was used and only $\Delta{m} >278\,$neV was considered. 

We designed an experimental setup, described below, to access this parameter space by leveraging higher magnetic fields and the resonance effect, and calculated the evolution for positive $\Delta{m}$ and $\theta_{0}$ down to $10^{-5}$. Since the neutron beam is unpolarized, the evolution calculation was performed for both polarizations and the results were averaged. Averaging over the velocity spectrum also effectively averages over the initial phases of the oscillation.

The experiment was performed at the SNS utilizing available equipment and leveraging the Magnetism Reflectometer instrument~\citep{doi:10.1080/10448630802210537,LAUTER20092543} (which primarily focuses on research of magnetic materials). This approach uses an interaction with a strongly neutron absorbing material to collapse the oscillating $(n,n')$ system in the magnetic field, instead of relying on the comparatively rare process of neutron $\beta$-decay. In this experimental setup, the cold neutron beam passes through a Cryomagnetics, Inc.\ superconducting split pair magnet~\cite{Cryomag5}, composed of two solenoids. The magnet has a $4.8\,$T magnetic field in the center, and peak magnetic field of $6.6\,$T at $6.3\,$cm upstream and downstream of magnet's center. A sintered boron carbide (B$_4$C) neutron absorber was placed in the center of the magnet. This absorber is transparent to mirror neutrons, and the nominal transmission of neutrons through the absorber assuming full density, no heterogeneity, and natural enrichment is $\sim10^{-12}$, estimated using PHITS~\cite{phits}. 

\begin{figure}
\centering
\includegraphics[width=0.475\textwidth]{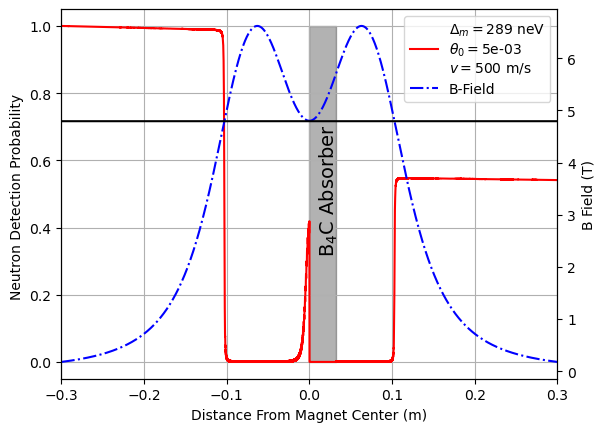}
\caption{\label{fig:Fig1}Evolution of single neutron through experimental setup. Description in the text. Color online.}
\end{figure}

Fig.~\ref{fig:Fig1} depicts an example of an individual neutron evolution (solid red line) through the experimental setup including magnetic field profile (dashed blue line) and absorber (grey fill), with $v=500\,$m/s and without phase averaging, and assuming $\Delta{m}=289\,$neV (corresponding to the central dip in magnetic field) and $\theta_0=5\times10^{-3}$, consistent with the neutron lifetime anomaly. Under these conditions, neutrons in the initial state $n$ would be nearly fully converted to the $n'$ state via the non-adiabatic LZ transition as they entered the magnet at 4.8\,T at -0.1\,m. Due to the split pair configuration of the magnet used and the particular choice of $\Delta{m}$ there is another LZ transition at 4.8\,T at the magnet center, and 60\% of neutrons would be in the $n'$ state on reaching the beam-catcher. While the $n'$ might naively be expected to be completely sterile in matter, we note that with the full treatment of the density matrix evolution, there is some attenuation in the absorber. After the absorber, the 55\% of neutrons remaining (in the $n'$ state) would be nearly fully converted as they exited the magnet at 4.8\,T at $+0.1$\,m. Reducing $\theta_0$ reduces the probability of conversion at each LZ transition, while the number of LZ transitions (0, 2, 3, or 4) depends on the value of $\Delta{m}$, indicated by where the horizontal black line crosses the blue dashed line in Fig.~\ref{fig:Fig1}. Evidence of neutrons regenerated from mirror neutrons would appear as a difference in detector counts above backgrounds between when the magnetic field is present and absent.
In the absence of $n\rightarrow n'$ oscillations, no signal above background would be detected. 

\begin{figure*}[t]
\centering
\includegraphics[width=\textwidth]{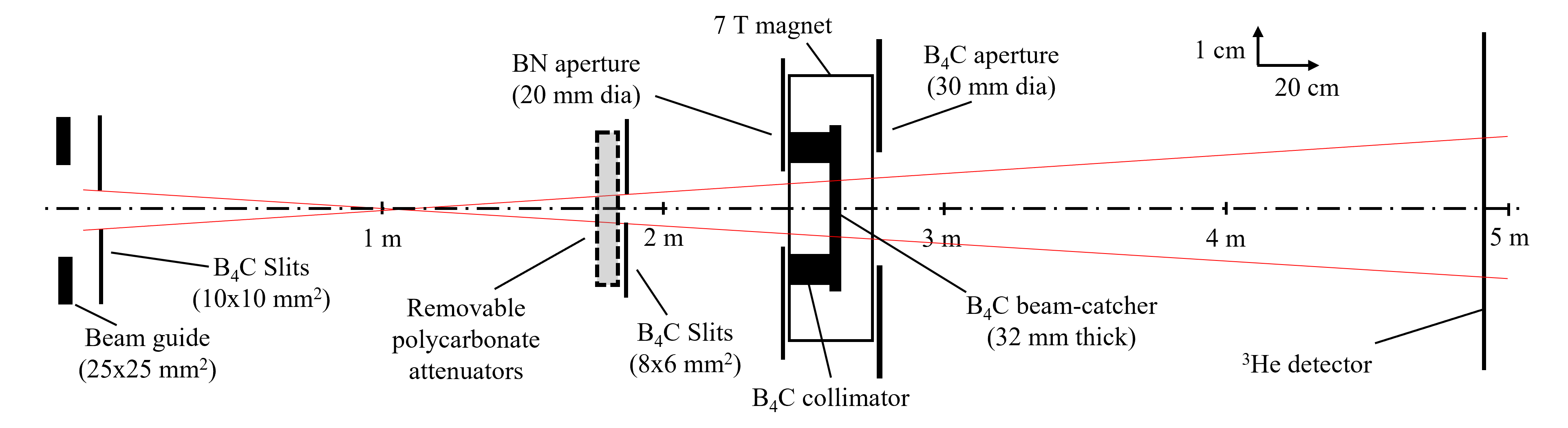}
\caption{\label{fig:Fig2}Schematic of the experiment at beamline 4A at SNS.  Description in the text.} 
\end{figure*}

The SNS operates at a nominal time-averaged proton power of 1.4\,MW with a repetition rate of 60\,Hz. The charge per proton pulse is measured with an accuracy of 3\%~\citep{Blokland:2006dh} and is included in the instrument data stream. The time-of-flight (TOF) between the short proton pulse and the signal in the neutron detector is used to determine the individual neutrons' velocity and wavelength. Three bandwidth limiting beam choppers prevent neutrons outside the chosen wavelength band from passing and prevent frame overlap. Neutrons were selected within a wavelength interval of $2.2$-$5.1\,${\AA} with a Center WaveLength (CWL) of $3.25\,${\AA}. The wavelength band was chosen to coincide with the maximum intensity of the spectrum available to the instrument.

A schematic of the experiment is shown in Fig.~\ref{fig:Fig2}. The cold neutron beam enters the experimental hall from the left of the diagram. After exiting the beam guide, the beam is collimated by a series of B$_4$C slits which define the extent and divergence of the beam. A removable assembly of stacked commercial polycarbonate (PC) $(C_{16}O_{3}H_{14})_{n}$ plates was installed for measurements requiring neutron attenuation.

Fig.~\ref{fig:Fig1} shows the calculated magnetic field map provided by the manufacturer and the extent of the B$_4$C absorber installed in the center of the magnet. The absorber consisted of a long B$_{4}$C cylinder with a thick B$_{4}$C plate end-cap glued to the downstream end, to serve as a ``beam-catcher''. A boron nitride aperature was installed at the front face of the magnet and a B$_4$C aperture was installed on the downstream face of the magnet to prevent scattering outside of the beam-catcher. 

Downstream of the magnet, the beam passed through a series of large apertures which reduce backgrounds from neutrons scattering in the room before encountering the instrument's neutron detector. The $^{3}$He gas-filled proportional-mode neutron detector resides within a movable shielded box. The detection efficiency is close to 90\%, slightly varying with the neutron wavelength. Each neutron count is recorded as an event containing time of flight to, and pixel position of, the detection event. 

Measurements were taken during a first allocation of beam time in July 2019 with the SNS operating at a proton power of 1.05\,MW, devoted to characterizing the neutron beam, and a second beam allocation in August 2019 with the SNS operating at full 1.4\,MW power, which included further characterizations and a search for the $n\rightarrow n'$ effect. The neutron intensity was normalized to the proton charge per frame (i.e. proton pulse window). At constant SNS proton power, the measured proton charge per frame was stable with a variation of about 0.7\% full width half-maximum. Time periods with beam instability and frames with proton charge outside of 1\% of the mean value were rejected. An additional cut in the TOF spectrum excluded direct fast neutron events arriving at the time of the prompt proton flash.

The neutron beam profile and TOF spectrum as measured by the main detector was found to be consistent within statistics between the July and August datasets, except for a small offset due to repositioning of the detector. The magnetic field was nearly constant inside the 1\,cm inner diameter of the beam-catcher, such that the beam extent and divergence inside the magnet had a small impact on the calculated conversion probability.

Since the maximum permitted neutron counting rate of the detector of 2~kcps is substantially lower than the total beam intensity, and as the detector electronics has a dead-time of 4\,$\mu{s}$ for each of the 256 vertical channels, the normalization to the neutron beam intensity was performed indirectly. We used the set of PC plates to vary the attenuation of the beam and extrapolate to zero attenuation. The neutron intensity was recorded using the main detector with 18, 20, and 24 PC plates in the July dataset. The 18 PC intensity was remeasured several times over the course of the run, including disassembly and reassembly of the stacked assembly, and the normalization was found to be stable to within $<0.5\%$. As the attenuation is mostly due to elastic scattering by hydrogen, additional scattering background was generated dependent on the number of PC plates. A 2D fit of the background to different functional forms was performed outside the beam signal region of interest (ROI) and the integral of the background function in that ROI was subtracted from the total intensity. Data were taken without attenuators but using pinholes in a Cd absorber placed 2.5\,m upstream of the detector to more accurately determine the unattenuated TOF spectrum and therefore the velocity distribution, needed for calculating the LZ transition probability. 

By extrapolating to the intensity with zero attenuation assuming an exponential dependence, we obtained the expected intensity incident on the B$_4$C beam-catcher, normalized to the integrated proton beam current. The reconstructed neutron beam intensity with no attenuation for the July 2019 measurements was found to be $(1.35\pm0.31)\times10^9$\,n/C between 2.2\,\AA{} and 5.1\,\AA. The uncertainty is dominated by the systematic error from different treatments of the scattering background subtraction. This extrapolation also indicates an attenuation of $\eta=0.675\pm0.008$ for each PC plate, compared to  $\eta=0.679\pm0.004$ calculated assuming cross sections from~\citep{NISTxs}, consistent with the  value estimated by the attenuation measurement.

The intensity calibration was performed in July 2019 at 1.05\,MW proton power; however, the $n\rightarrow n'$ measurements were performed in August 2019 at a proton power of 1.4\,MW.  The neutron beam intensity per unit proton charge can vary depending on the properties of the neutron moderator, which is the source of the cold neutron beam, and the repeatability of the slit configurations. Therefore, our estimated intensity normalization was reduced by $78\pm2\%$ to $(1.05\pm0.31)\times10^9$\,n/C, due to the lower measured neutron intensity per MW.

To search for evidence of the $n\rightarrow n'$ transformation, we installed the B$_{4}$C beam-catcher in the magnet such that the full intensity of the neutron beam was totally absorbed. Any mirror neutrons generated in the magnetic field before the beam-catcher passed through the $B_{4}C$ and were regenerated into neutrons in the magnetic field after the absorber to be counted in the detector within the previously determined signal ROI. The polarity of the magnetic field does not change the magnitude of the effect for unpolarized neutrons since both polarizations were averaged (equation 14 of \citep{Berezhiani:2018eds}). Data were also taken at zero magnetic field to reduce the effect and served as a no-signal comparison. Backgrounds were dominated by neutrons scattered in the room due to incomplete shielding of the detector.

\begin{table}
\centering
\begin{tabular}{rcccc}
%\hline
B Field & ROI & Raw Counts & Charge & Counts/C \\
\hline
$-4.8$\,T &sig& $4976\pm70$ & 8.8\,C & $564.6\pm 8.0$ \\
$+4.8$\,T &sig& $7748\pm88$ & 13.8\,C & $561.0\pm 6.4$ \\
$0$\,T &sig& $6631\pm81$ & 11.9\,C & $558.2\pm 6.9$ \\
$0$\,T &bkg& $6387\pm80$ & 11.9\,C & $547\pm16^*$\\
%\hline
\end{tabular}
\caption{\label{tab:rates} Neutrons detected on the main detector at different magnetic field values in the signal or background ROI, normalized to integrated proton charge. Uncertainties are statistical and represent one standard deviation. The background ROI counts/C includes a $+1.8\pm2.8\%$ efficiency correction.}
\end{table}

Measurements with the B$_{4}$C beam-catcher blocking the neutron beam were performed in an alternating sequence of typically 1\,hour runs with a magnetic field at the B$_4$C (magnet center) of $+4.8\,$T, $0\,$T, and $-4.8\,$T. No evidence of transmission of the neutron beam through the beam-catcher was observed in any configuration. The total integrated counts within the background and signal ROI at different fields were statistically equivalent (Table~\ref{tab:rates}). 

The apparent transmission $p$ through the beam-catcher due to neutron oscillations was taken from the difference in the total integrated counts in the ROI of $(562.4\pm5.0)$\,n/C with magnetic field (average of both polarities) and of $(558.2\pm6.9)$\,n/C without magnetic field to estimate an apparent signal of $(4.2\pm8.5)$\,n/C. The difference is then divided by the incident neutron intensity of $(1.05\pm0.31)\times10^9$\,n/C, as described above, giving an apparent transmission of $(0.4\pm1.2)\times10^{-8}$. We used the Feldman-Cousins method~\citep{PhysRevD.57.3873} to determine a $95\%$ confidence limit (C.L.) on the apparent transmission of $p<2.5\times10^{-8}$. 

\begin{figure}[t]
\centering
\includegraphics[width=0.475\textwidth]{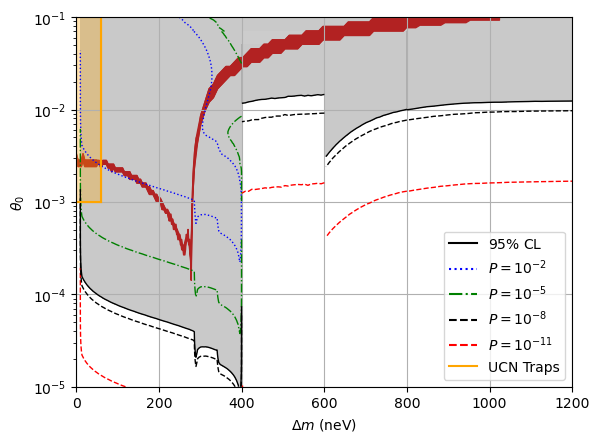}
\caption{\label{fig:Fig3}The parameter space of the mirror matter model~\citep{Berezhiani:2018eds} excluded with 95\%\,C.L. (gray region) and prediction consistent with the neutron lifetime anomaly (red band). Calculated probability of transmission is given by the dashed contour lines. The region 400\,neV to 600\,neV uses a different dataset. Color online.} 
\end{figure}

The calculated probability of transmission for the oscillation process as described above is shown by the dashed contour lines in Fig.~\ref{fig:Fig3}. The Feldman-Cousins upper limit that we measured separates Fig.~\ref{fig:Fig3} into the portion of $\Delta{m}$, $\theta_0$ space excluded by our measurement (filled gray), where $p>2.5\times10^{-8}$, from regions where we do not have statistical sensitivity (white). The features at 289\,neV, 340\,neV, and 400\,neV correspond to the central dip, inflection points, and maximum, respectively, of the magnetic field in Fig.~\ref{fig:Fig1}.  The complex optical potential of the strongly absorbing B$_4$C together with the $4.8$\,T magnetic field creates a strong absorption resonance that reduces our sensitivity between 400\,neV and 600\,neV~\citep{Kamyshkov:2021kzi}. In this region, therefore, the normalized counts in the signal and background ROIs at B=0\,T were compared using the same procedure to obtain $p<5.5\times10^{-8}$ (95\% C.L.) due to field-free adiabatic transitions, and the calculated probability at B=0 is used for exclusion. For $\Delta{m}$ much larger than the magnetic potential, transitions are adiabatic and the regeneration probability asymptotically approaches the magnetic field-free limit and becomes constant~\citep{Berezhiani:2018eds}.

The red band in Fig.~\ref{fig:Fig3} corresponds to the parameter space that would explain the neutron lifetime anomaly. The dips in the red band at $\sim$260\,neV and 278\,neV correspond to flatter regions in the Beam Lifetime magnetic field profile near $\sim4.3$\,T and $\sim4.6$\,T respectively~\cite{Nico:2004ie}. Above 278\,neV, $n\rightarrow {n'}$ transitions occur adiabatically. Uncertainties due to calculational and experimental inputs to the model are negligible.

In summary, we have conducted an experiment using the novel approach of cold neutron regeneration and excluded the non-degenerate mirror matter model as an explanation for the neutron lifetime anomaly for mass splittings above 10\,neV. We note that the attenuation of mirror neutrons in matter is an important consideration in calculating limits for this model. Limits from  an experimental study of anomalous losses per collision in UCN trap experiments \citep{Serebrov:2004zf}, reinterpreted as $n\rightarrow n'$ disappearance, exclude $\theta_0 \gtrsim 10^{-3}$ \citep{Berezhiani:2018eds} for mass splittings below $\sim60$\,neV (yellow box in Fig.~\ref{fig:Fig3}).  This result does not provide constraints on alternative models of $n\rightarrow {n'}$ oscillation, such as the minimal model where $\Delta{m}=0$ but a mirror magnetic field $B'$ is assumed~\citep{Berezhiani:2009ldq} or models with a common neutron transition magnetic moment (nTMM) for $n$ and $n'$ components~\citep{Berezhiani:2018qqw}. This measurement represents the first of a broad planned program using neutron scattering instruments at ORNL to search for processes violating baryon minus lepton number $\mathcal{B-L}$ and $\mathcal{B}$ by one unit, similar to the process $n\rightarrow\bar{n}$ which violates $\mathcal{B-L}$ and $\mathcal{B}$ by two units~\citep{Proceedings:2020nzz,Phillips:2014fgb}.

\begin{acknowledgments}
This research was sponsored by the U.S. Department of Energy (DOE), Office of Science, Office of Nuclear Physics [contract DE-AC05-00OR22725], by the Laboratory Directed Research and Development Program [project 8215] of Oak Ridge National Laboratory, managed by UT-Battelle, LLC, for the U. S. DOE, and in part by the U.S. DOE, Office of Science, Office of Workforce Development for Teachers and Scientists (WDTS) under the Science Undergraduate Laboratory Internship program. 
J. L. Barrow was supported by the U.S. DOE, Office of Science, Office of WDTS, Office of Science Graduate Student Research (SCGSR) program. The SCGSR program is administered by the Oak Ridge Institute for Science and Education for the DOE under Contract No. DESC0014664.
L. Varriano was supported by a National Science Foundation Graduate Research Fellowship under Grant No. DGE-1746045. This research used resources at the Spallation Neutron Source, a DOE Office of Science User Facility operated by the Oak Ridge National Laboratory. 
\end{acknowledgments}

\bibliographystyle{apsrev4-2}
\bibliography{refs-PRLstyle-revised}
%\bibliography{test}

\end{document}